\documentclass{emulateapj}

\usepackage{epsfig, natbib}
\usepackage{microtype}

\begin{document}

\title{Protostellar Disk Formation Enabled by Weak, Misaligned Magnetic Fields}

\author{Mark R.~Krumholz\altaffilmark{1}, Richard M.~Crutcher\altaffilmark{2}, \& Charles L.~H.~Hull\altaffilmark{3}}

\altaffiltext{1}{Department of Astronomy and Astrophysics,
         University of California, Santa Cruz, CA 95064 USA;
         mkrumhol@ucsc.edu}
\altaffiltext{2}{Department of Astronomy, University of Illinois at Urbana-Champaign, 1002 W Green Street, Urbana, IL 61801, USA}
\altaffiltext{3}{Astronomy Department \& Radio Astronomy Laboratory, University of California, Berkeley, CA 94720-3411 USA}

\begin{abstract}
The gas from which stars form is magnetized, and strong magnetic fields can efficiently transport angular momentum. Most theoretical models of this phenomenon find that it should prevent formation of large ($>100$ AU), rotationally-supported disks around most protostars, even when non-ideal magnetohydrodynamic (MHD) effects that allow the field and gas to decouple are taken into account. Using recent observations of magnetic field strengths and orientations in protostellar cores, we show that this conclusion is incorrect. The distribution of magnetic field strengths is very broad, and alignments between fields and angular momentum vectors within protostellar cores are essentially random. By combining the field strength and misalignment data with MHD simulations showing that disk formation is expected for both weak and misaligned fields, we show that these observations imply that we should expect disk fractions of $\sim$10 -- 50\% even when protostars are still deeply embedded in their parent cores, and even if the gas is governed by ideal MHD.
\end{abstract}

\keywords{magnetic fields --- protoplanetary disks --- stars: formation --- stars: protostars --- stars: winds, outflows}

\section{Introduction}

Stars form via the collapse of dense, magnetized cores of interstellar gas. As the gas contracts, it spins up to conserve angular momentum and twists magnetic field lines. This launches torsional Alfv\'{e}n waves that transport angular momentum away from the densest, collapsing region. In cores that have aligned field and rotation axes, and flux-to-mass ratios within a factor of a few of the critical value (defined as the value at which the magnetic field is able to prevent collapse entirely), this magnetic braking is so effective that Keplerian disks do not form \citep{galli06a, price07a, mellon08a,  hennebelle09b, krasnopolsky12a}. For fields that are stronger (but that are still weak enough to render the core magnetically supercritical), the outcome is no disk at all; for weaker fields, the result is a sub-Keplerian ``pseudodisk'' supported by magnetic pressure rather than rotation.

Since Keplerian disks and the planetary systems they produce are observed to be ubiquitous around optically-visible young stars \citep[\textit{e.g.,}][]{haisch01a}, this result presents an obvious problem. A number of possible solutions have been proposed. One is that disks do not appear until the majority of the surrounding protostellar core has been accreted, at which point the inertia of the envelope into which the twisted fields deposit angular momentum is greatly reduced, preventing efficient magnetic braking \citep{mellon09a, krasnopolsky12a}. In this scenario, disks should not appear until the Class II phase, when the envelope is cleared and protostars become optically visible. This proposal faces two severe challenges, however.  First, observations now directly demonstrate that extended disks, at least some of them Keplerian, are present even in Class 0 and Class I sources with large envelopes \citep{jorgensen09a, enoch09a, takakuwa12a, tobin12a}.\footnote{Note that \citet{maury10a} have searched for fragmentation or other circumstellar structures around class 0 sources on few hundred AU scales, and report a null result. However, due to their interferometer settings, they are insensitive to disks with masses of $\sim 0.1$ $M_\odot$ or less.} And second, even if the gas in a protostellar core has been drained, the magnetic field lines threading the pseudo-disk still connect to the much larger surrounding molecular cloud, and it is unclear why magnetic braking would cease to operate. 

Another possible solution is that non-ideal MHD effects allow the magnetic field to decouple from the gas, preventing efficient magnetic braking. A number of authors have investigated whether ion-neutral drift \citep{mellon09a, duffin09a, li11b} or Ohmic dissipation \citep{dapp10a, machida11a, li11b} might allow Keplerian disks to form, and have found that either they fail to produce disks at all, or they produce disks that are only $\sim$10 AU in size. In contrast, the observed disks even around deeply embedded protostars are an order of magnitude larger \citep{jorgensen09a, enoch09a, takakuwa12a, tobin12a}. Turbulent reconnection might also allow disk formation \citep{santos-lima12a}, but thus far this has been demonstrated to be effective only in the presence of supersonic turbulence. In contrast, observations \citep{kirk07a, andre07a, rosolowsky08a, pineda10a} and theory \citep{offner08a} show that typical low-mass stellar cores are at most trans-sonically turbulent.

A third option, which we investigate here, is that disk formation might be enabled by a combination of low magnetic field strengths and misalignment between the fields and the angular momentum vector of the protostellar cores they thread. Simulations of misaligned cores indicate that extended Keplerian disks can form if the field is weak enough and the angle between the field and the rotation axis is large enough \citep{hennebelle09b, ciardi10a, joos12a}. While this possibility is promising, up until now it has not been possible to determine if this is a viable option for disk formation in general, because the distribution of core magnetic field strengths and field/rotation misalignments was unknown. However, recent observations have changed this situation, providing good statistical measures of both quantities.

In the following sections we first discuss the observational evidence regarding magnetic field strengths, and then that regarding field/rotation alignment. We then combine these observations with simulations in order to calculate disk fractions. Finally, we discuss the implications of our results.

\section{Observational Inputs}

\subsection{Magnetic Field Strengths in Protostellar Cores}

Field strengths can be measured via the Zeeman effect \citep{crutcher10a}, although with current sensitivities it is only possible to measure the line-of-sight component $B_{\rm los}$ of the magnetic vector \textbf{B}. Hence, only lower limits on the total magnetic field strength $B_{\rm total}$, and therefore lower limits on the magnetic flux-to-mass ratio $\Phi/M$, may be measured directly. If gravity is balanced by magnetic support, $\Phi/M$ is said to be critical. Hereafter all $\Phi/M$ values are normalized by the critical value, so if $\Phi/M < 1$ for a cloud it is supercritical and the magnetic field is too weak to support the cloud against gravity. For a sufficiently large sample of clouds one would expect some {\bf B} to be approximately along the line of sight, so if subcritical clouds existed some observed $\Phi/M$ lower limits would be above the critical value. However, evaluation of available surveys of Zeeman observations of OH and CN lines \citep{crutcher12a} has shown that there are no directly measured subcritical $\Phi/M$ in molecular clouds; that is, for all Zeeman measurements of $B_{\rm los}$, $\Phi/M < 1$. 


For a sample of Zeeman observations that includes both detections and non-detections, one obtains the PDF (probability density function) of $B_{\rm los}$, $P(B_{\rm los})$. There are two factors that determine $P(B_{\rm los})$. One is the distribution of angles between the direction of {\bf B} and the line of sight along which $B_{\rm los}$ is measured. For a sufficiently large sample of clouds, one assumes that the angles are randomly distributed. The other factor is $P(B_{\rm total})$. Regardless of the functional form of $P(B_{\rm total})$, the observed distribution $P(B_{\rm los})$ implies that the mean value of $B_{\rm total}$ in a sample is approximately 2 times the mean value of the $B_{\rm los}$ \citep{heiles05b}. (A delta-function form for $P(B_{\rm total})$ gives $\langle{B}_{\rm total}\rangle = 2.0 \times \langle \vert B_{\rm los}\vert \rangle$, and a flat $P(B_{\rm total})$ gives $\langle{B}_{\rm total}\rangle = 1.9 \times \langle \vert B_{\rm los}\vert \rangle$.) Hence, the most straightforward analysis of a set of Zeeman measurements is to compute the magnitude of the mean of the $B_{\rm total}$ as $\langle{B}_{\rm total}\rangle = 2 \times \langle \vert B_{\rm los}\vert \rangle$. Note that $B_{\rm los}$ is signed, with the sign indicating whether {\bf B} points toward or away from the observer.

A more sophisticated Bayesian analysis of the observed distribution of $P(B_{\rm los})$ further indicates that the most probable functional form for $P(B_{\rm total})$ is flat \citep{crutcher10a}; that is, at any density $n_{\rm H}$ a sample of molecular clouds would have $B_{\rm total}$ ranging with equal probability from near zero up to a maximum $B_{\rm max}$, with that maximum scaling as a power law in $n_{\rm H}$. \citet{crutcher12a} reviewed studies of $\langle\Phi/M\rangle$ in a large sample of molecular clouds and concluded that the mean flux-to-mass ratio $\langle\Phi/M\rangle \approx 0.5$. Combined with the results that $P(B_{\rm total})$ is flat, this implies that the most likely true distribution of $\Phi/M$ values is flat, with a minimum of 0 and a maximum of $\Phi/M=1$. 

\subsection{Field/Rotation Misalignment}

The distribution of magnetic field strengths provides one of the two observational inputs required to calculate the disk fraction. The other is the distribution of relative angles between core magnetic fields and angular momentum vectors. The direction of a core's angular momentum vector can be inferred from the orientation of its outflow, since outflows are launched perpendicular to circumstellar disks. Field directions in the plane of the sky can be measured via dust polarization. Under almost all circumstances, spinning dust grains are expected to align themselves with their long axes perpendicular to the magnetic field \citep{lazarian07a}, leading to both polarized emission and absorption. In dense cores, which are opaque to background starlight, mapping the polarized thermal emission from magnetically aligned dust grains is the most practical means of studying the field morphology on small scales.

Large-scale sub-mm dust polarization maps with resolutions of $\sim$20$^{\prime\prime}$ have been obtained toward many sources, but significantly higher angular resolution is required to study the field geometry in the densest regions of protostellar cores; this requires interferometric polarization maps. Until recently only about a dozen sources had been mapped (\textit{e.g.,}~NGC~1333-IRAS~4A: \citealt{girart06a}; IRAS~16293: \citealt{rao09a}; and Orion~KL: \citealt{rao98a, tang10a}), which was too few to allow statistical analysis. However, the 1~mm dual-polarization receiver system at CARMA (the Combined Array for Research in Millimeter-wave Astronomy) has enabled high-resolution mapping of dust polarization toward many more sources. \citet{hull12a} present results from the TADPOL survey\footnote{{\bf T}elescope {\bf A}rray {\bf D}oing {\bf POL}arization \\ \url{http://tadpol.astro.illinois.edu/}}, a CARMA key project. They focus on nearby, low-mass protostars that have clear outflow and magnetic field orientations.

The TADPOL results strongly rule out a scenario where magnetic fields and disks are aligned. Rather, their results are consistent with a random distribution of orientations between the field and the outflow. The data are even consistent with a preferential anti-alignment between outflows and magnetic fields.  If one assumes that outflows emerge along the rotation axes of circumstellar disks, as is expected in magnetocentrifugally launched wind model \citep{pudritz86a, shu94a}, then the results imply that these disks are not aligned with the fields in the cores from which they formed, or that they are preferentially anti-aligned.

\section{The Protostellar Disk Fraction}

\begin{figure}
\epsscale{1.2}
\plotone{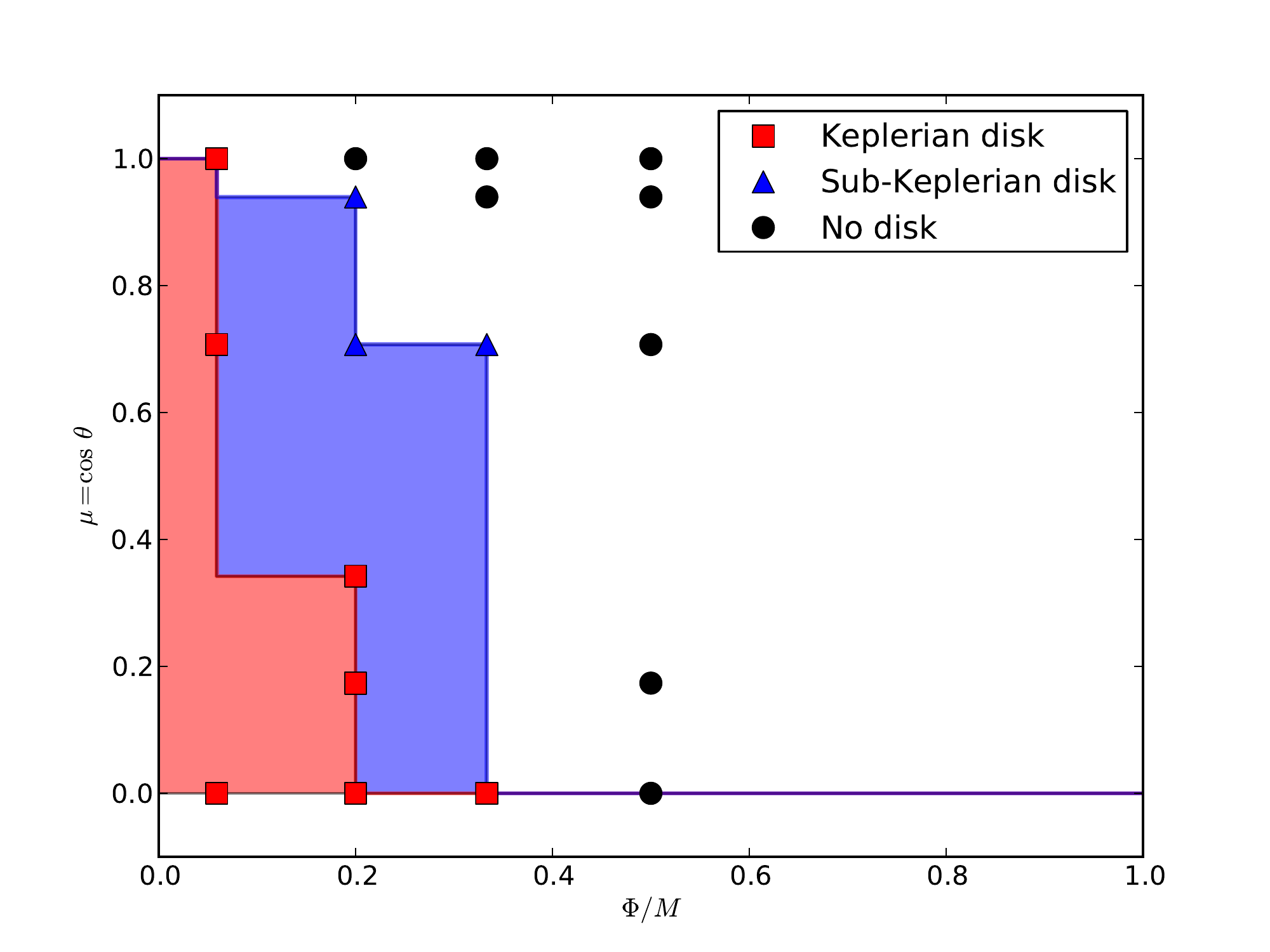}
\caption{
\label{fig:results1}
The parameter space of $(\Phi/M, \mu)$ that describes the strength of core magnetic fields, and their orientation relative to the angular momentum vectors; $\mu$ is the cosine of the misalignment angle. Observations indicate that cores are uniformly distributed in the range $\Phi/M = 0 - 1$, and $\mu = 0-1$. Points show simulation results \citep{joos12a}, indicating whether a simulation with those parameters formed no disk, a sub-Keplerian disk, or a Keplerian disk. The shaded regions show the range in parameter space over which Keplerian disks (red) and non-Keplerian disks (blue) form, under our minimal assumptions. The fraction of cores that will produce Keplerian disks is the area of the red region, and the fraction that will produce any disk at all is the sum of the areas of the red and blue regions.\\
}
\end{figure}

\begin{figure}
\epsscale{1.2}
\plotone{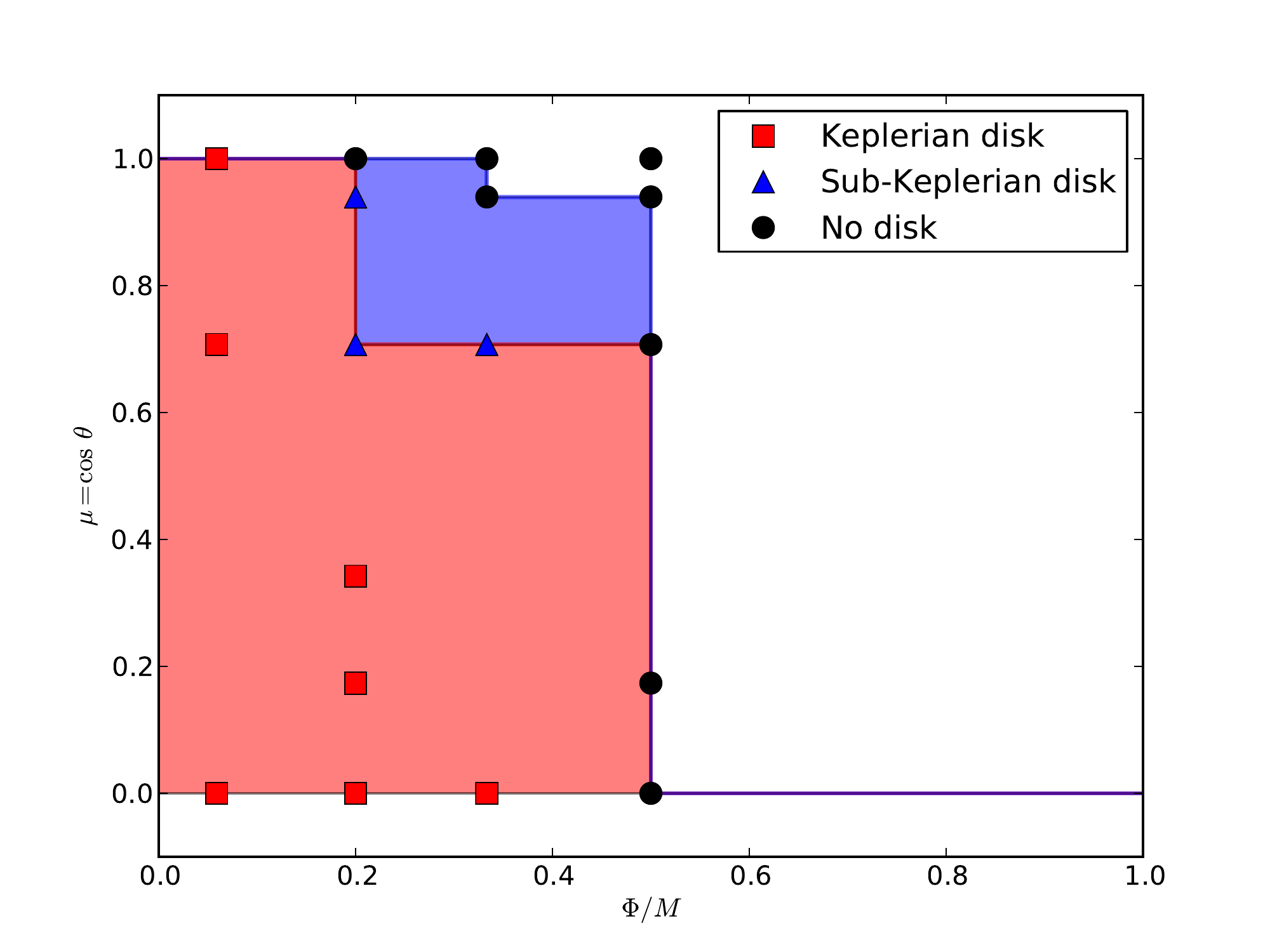}
\caption{
\label{fig:results2}
Same as Figure \ref{fig:results1}, but now the shaded regions show maximal rather than minimal assumptions about where in parameter space disks form.\\
}
\end{figure}

The final input required to calculate the disk fraction is a calculation of where disks are expected to form in the parameter space of magnetic field strength and field/rotation misalignment angle. The most complete numerical study to date \citep{joos12a} includes 18 simulations using ideal MHD that sample values of $\Phi/M$ in the range $0.06-0.5$, and misalignment angles $\theta$ from $0-90^\circ$. Figures \ref{fig:results1} and \ref{fig:results2} summarize the numerical results, where each result is classified as producing a Keplerian disk ($\sim$100 AU in size or larger), a sub-Keplerian disk (again at least $\sim$100 AU in size), or no disk at all.

The observed distribution of field strength implies that cores are uniformly distributed in $\Phi/M$ from 0 to 1. If we adopt the more conservative TADPOL result that field/rotation alignment is random, as opposed to preferentially anti-alignned, then cores are also uniformly distributed in $\mu = \cos\theta$ from 0 to 1, implying that the distribution in the $(\Phi/M, \mu)$ plane is uniform from 0 to 1 in both dimensions. In this case the fraction of systems that possess disks, Keplerian or otherwise, is simply the fraction of the area of the unit square in the $(\Phi/M, \mu)$ plane over which disks form. To estimate this, we can make either minimal or maximal assumptions about disk formation based on the simulations. In the minimal case we assume that if a disk forms in a simulation at a point $(\Phi/M, \mu)_1$ in parameter space, but not at a point $(\Phi/M,\mu)_2$ where either $(\Phi/M)_2 > (\Phi/M)_1$ or $\mu_2>\mu_1$, then disks form only at $\Phi/M \leq (\Phi/M)_1$ and $\mu \leq \mu_1$. A maximal assumption is the opposite: disks form for all $\Phi/M < (\Phi/M)_2$ and $\mu < \mu_2$. The shaded regions in Figures \ref{fig:results1} and \ref{fig:results2} summarize the two cases. With either assumption we can compute the disk fraction simply by integrating over the shaded regions. Doing so we find that Keplerian disks are expected to form in $(11\%, 48\%)$ of cores, and either Keplerian or non-rotationally-supported pseudo-disks in $(29\%, 50\%)$ of cores, where the first number corresponds to the minimal assumption, and the second to the maximal.

\section{Discussion and Conclusions}

Our calculation that $10-50\%$ of cores should produce Keplerian disks provides a natural explanation for the significant number of Class 0 and I sources around which disks have been observed. At present we lack a full census of the disk fraction at these early stages, and thus it is unclear if the percentages we compute are consistent with the actual fraction of such systems that have disks, or if some other mechanism will be required.

While our results help alleviate the problem at the Class 0 and I stage, a disk fraction of 50\% still too small in comparison to what is observed at the Class II stage, where the disk fraction approaches unity \citep{haisch01a}. The question, then, is what mechanism might explain such a high disk fraction at these later stages. One possibility is that we have been too conservative in adopting a uniform distribution of field/rotation orientations, and that in fact fields are preferentially anti-aligned with core angular momentum vectors. If this is the case, then cores will preferentially occupy the region near $\mu = 0$, and the disk fraction will be higher than we have estimated. This possibility may be checked by further polarization measurements of the sort performed by \citet{hull12a}.

A second possible explanation for the high fraction of disks in Class II sources is that it is the result of a combination of misalignment with non-ideal MHD effects. The values illustrated in Figures \ref{fig:results1} and \ref{fig:results2} do not include any non-ideal MHD effects, such as ion-neutral drift, Ohmic dissipation, or turbulent reconnection. Any of these effects would probably enhance the ability of disks to form, since they would reduce the ability of magnetic fields to extract angular momentum from infalling gas.

A final possibility is that a reduction in the inertia of the envelope might yield a Class II disk fraction that exceeds that found in Class 0 and Class I sources. In this case the disk fraction would only be $\sim$$10-50\%$ during the Class 0/I phase, but would rise to nearly 100\% at the transition to the Class II phase as the envelope depletes. One may distinguish between this scenario and the previous two by measuring the disk fraction for Class 0/I systems. If the results are $\sim$$10-50\%$, that is consistent with non-ideal effects being unimportant and with a uniform distribution of disk-rotation angles, in which case the entire difference between Class 0/I and Class II sources would arise from a reduction in envelope inertia. On the other hand, a disk fraction near unity for Class 0/I sources would favor either non-ideal effects or preferential field/rotation anti-alignment as an explanation.

\acknowledgements M.R.K.~acknowledges support from the Alfred P.~Sloan Foundation, NSF CAREER grant AST-0955300, and NASA through a Chandra Space Telescope Grant. R.M.C.~acknowledges support from grant NSF AST 10-07713 and the CARMA project. C.L.L.H. acknowledges the support from an NSF Graduate Fellowship. Support for CARMA construction was derived from the states of California, Illinois, and Maryland, the James S. McDonnell Foundation, the Gordon and Betty Moore Foundation, the Kenneth T. and Eileen L. Norris Foundation, the University of Chicago, the Associates of the California Institute of Technology, and the National Science Foundation. Ongoing CARMA development and operations are supported by the National Science Foundation under a cooperative agreement, and by the CARMA partner universities.


%


\end{document}